\documentclass[reprint,aps,prb,longbibliography,amsmath,amssymb,superscriptaddress]{revtex4-2}
\usepackage{booktabs}
\usepackage{hhline}
\usepackage{amsmath}
\usepackage{amssymb}
\usepackage[version=4]{mhchem}
\usepackage{siunitx}
\DeclareSIUnit\angstrom{\text{\AA}}

\usepackage{svg}
\usepackage{graphicx}
\usepackage{dcolumn}
\usepackage{bm}

\usepackage{hyperref}
\hypersetup{
unicode=false, pdftoolbar=true, pdfmenubar=true, pdffitwindow=false, pdfstartview={FitH}, colorlinks=true, linkcolor=blue, citecolor=blue, urlcolor=blue, }

\usepackage{etoolbox}

\graphicspath{ {./images/}}
\usepackage{multirow}

\pdfoutput=1

\begin{document}

\title{Real-space electrostatic sums through truncation error analysis}

\newcommand\HARVARD{John A. Paulson School of Engineering and Applied Sciences,
Harvard University, Cambridge, MA, USA}

\author{Chuin Wei Tan}
\thanks{Corresponding author}
\thanks{\href{mailto:chuinwei_tan@g.harvard.edu}{chuinwei\_tan@g.harvard.edu}}
\affiliation{\HARVARD}

\author{William C. Witt}
\affiliation{\HARVARD}

\author{Boris Kozinsky}
\thanks{Corresponding author}
\thanks{\href{mailto:bkoz@g.harvard.edu}{bkoz@g.harvard.edu}}
\affiliation{\HARVARD}
\affiliation{Robert Bosch LLC Research and Technology Center, Watertown, MA, USA}

\begin{abstract}
\noindent
Methods that treat electrostatics in real space, without explicit reciprocal-space evaluation, are appealing for large-scale atomistic simulations.
They typically employ local charge-neutralization schemes, but their relationship to the underlying error contributions of the full Ewald sum has not been fully characterized.
A principled basis for parameter selection is also lacking.
In this work, we show that the correction terms of Wolf's and Pickard's real-space methods approximate the truncation error of real-space electrostatic sums.
This perspective motivates simplified fixed-radii and force-shifted Pickard variants that preserve the truncation-error compensation behavior while enforcing continuity of pair energy and force at the cutoff.
We derive a nonempirical scaling relation for the optimal path in the parameter space of the cutoff radius and screening distance, generalizing Pickard's empirical heuristic and explaining when and why it can fail.
A systematic scheme for restoring omitted reciprocal contributions is introduced, which accelerates convergence to the full Ewald energy.
\end{abstract}

\maketitle

\section{Introduction}
The electrostatic interaction of point charges is central to the quantitative description of materials.
For \(N\) point charges, its governing expression is the Coulomb pair sum,
\begin{equation} \label{eq:coulomb_pair_sum}
    E = \frac12 \sum_{i=1}^N \sum_{j\ne i}^N\frac{q_i q_j}{r_{ij}},
\end{equation}
where \(q_i\) is the charge on site \(i\) and \(r_{ij}\) is the distance between sites \(i\) and \(j\).
Under periodic boundary conditions, the Ewald sum provides a formally exact decomposition of Eq.~\eqref{eq:coulomb_pair_sum} into real-space and reciprocal-space contributions that both converge rapidly, with a screening distance controlling the balance between them~\citep{ewald1921berechnung}.
However, the reciprocal-space term is intrinsically nonlocal and poses difficulties for large-scale distributed simulations, where global communication and synchronization can dominate the cost.
Real-space alternatives that approximate Coulomb interactions locally without explicit reciprocal-space evaluation are thus attractive.
Ideally, such a scheme should be maximally accurate at a given real-space cutoff radius, converge systematically as the cutoff increases, and produce continuous forces, stresses, and higher derivatives where needed.

The standard entry point in real-space electrostatics is the Wolf sum, in which the net charge enclosed within each cutoff sphere is neutralized by a compensating charge placed on the cutoff surface~\citep{wolf1999exact,zahn2002enhancement,fennell2006ewald}.
This construction has since been extended in multiple directions~\citep{fukuda2011molecular,fanourgakis2015extension,waibel2018modification,waibel2018modified,stenqvist2020generalized}.
A different route was taken by Pickard, who proposed a local neutralization scheme in which the compensating charge is distributed continuously throughout space~\citep{PhysRevMaterials.2.013806}.
These charge-neutralization constructions regularize and accelerate convergence of the finite-cutoff real-space sum.
Despite their success~\citep{fennell2006ewald,PhysRevMaterials.2.013806}, the relationship between these corrections and the underlying Ewald sum has not been fully characterized.
Furthermore, these methods depend on two parameters, the cutoff radius and the screening distance, whose optimal choice has been deemed ``arbitrary''~\citep{hansen2012simplistic}, with ``no clear criterion for choosing these parameters''~\citep{hens2017molecular}.
They are typically determined by empirical heuristics~\citep{demontis2001application,zahn2002enhancement,hansen2012simplistic,hens2017molecular,PhysRevMaterials.2.013806}, but the absence of a theoretical basis makes it unclear when and why such heuristics can fail.

In this work, we show that Wolf's and Pickard's local charge compensation terms approximate the real-space truncation error of the Ewald sum.
Guided by this perspective, we introduce a simplified fixed-radii variant and a force-shifted extension that enforces continuity of both pair energy and force at the real-space cutoff.
We then derive a nonempirical scaling relation for the optimal parameters of damped real-space methods, which explains Pickard's empirical heuristic and extends it beyond its original scope.
We also show that omitted reciprocal contributions can be restored systematically to accelerate convergence.

Due to the numerically sensitive nature of this work, we evaluate all lattice sums in 80-bit extended precision with a 64-bit significand.

\section{Ewald Sum}

The Ewald energy~\citep{ewald1921berechnung} per unit cell is
\begin{equation}
    E_{\mathrm{Ewald}} = E_{\mathrm{real}}^{\infty} + E_{\mathrm{recip}}^{\infty} + E_{\mathrm{self}} + E_{\mathrm{bg}},
\label{eq:ewald_sum}
\end{equation}
where the terms are defined below.
The short-range (real-space) contribution
\begin{equation}
    E_{\mathrm{real}}^{\infty}
    =\lim_{R_c\to\infty}
    \frac12 \sum_{i=1}^{N_\text{cell}} \sum_{\substack{j\ne i\\ r_{ij}\le R_c}}
    q_i q_j\,\phi_{R_d}(r_{ij})
\end{equation}
accounts for damped Coulomb interactions via the kernel
\begin{equation}
    \phi_{R_d}(r)
    =
    \frac{\operatorname{erfc}(\alpha r)}{r}
    =
    \frac{\operatorname{erfc}(r / R_d)}{r}
    ,
\label{eq:phi_def}
\end{equation}
which is typically parameterized by a screening parameter \(\alpha\), or, equivalently, a screening distance \(R_d = 1/\alpha\).
The sum over \(i\) runs over the \(N_\text{cell}\) charges in the unit cell, and the index set for \(j\) includes all periodic images of the charges.
The long-range (reciprocal-space) contribution is
\begin{equation}
    E_{\mathrm{recip}}^{\infty}
    =\lim_{G_c\to\infty}
    \frac{2\pi}{\Omega}\sum_{\substack{\mathbf{G}\neq \mathbf{0}\\ |\mathbf{G}|<G_c}}
    \frac{e^{-G^2/(4\alpha^2)}}{G^2}\,|S(\mathbf{G})|^2,
\label{eq:ewald_recip}
\end{equation}
where \(\Omega\) is the cell volume and
\begin{equation}
    S(\mathbf{G})=\sum_{j=1}^{N_\text{cell}} q_j e^{i\mathbf{G}\cdot\mathbf{r}_j}
\end{equation}
is the charge-weighted structure factor for nonzero reciprocal lattice vectors \(\mathbf{G}\).
Finally,
\begin{equation}
    E_{\mathrm{self}}=-\frac{\alpha}{\sqrt{\pi}}\sum_{i=1}^{N_\text{cell}} q_i^2
\end{equation}
is a self-interaction correction that removes the spurious interaction of each charge with its own screening cloud, and
\begin{equation}
    E_{\mathrm{bg}}=-\frac{\pi Q^2}{2\Omega\alpha^2}
\label{eq:ewald_bg}
\end{equation}
is a background-neutralization term. Here,
\begin{equation}
    Q=\sum_{i=1}^{N_\text{cell}} q_i
\end{equation}
is the total ionic charge in the unit cell.
The cutoffs \(R_c\) and \(G_c\) truncate the real-space and reciprocal-space sums in numerical computation and are taken to infinity in the exact expressions above.
In practice, one computes
\begin{equation}
    E_{\mathrm{Ewald}}^{R_c,G_c}
    =
    E_{\mathrm{real}}^{R_c}+
    E_{\mathrm{recip}}^{G_c}+
    E_{\mathrm{self}}+
    E_{\mathrm{bg}},
\end{equation}
where the finite-cutoff terms are
\begin{equation}\label{eq:ewald_real_rc}
    E_{\mathrm{real}}^{R_c}
    =\frac12\sum_{i=1}^{N_\text{cell}}\sum_{\substack{j\ne i\\ r_{ij}\le R_c}}
    q_i q_j\,\phi_{R_d}(r_{ij}),
\end{equation}
and
\begin{equation}
    E_{\mathrm{recip}}^{G_c}
    =\frac{2\pi}{\Omega}\sum_{\substack{\mathbf{G}\neq \mathbf{0}\\ |\mathbf{G}|<G_c}}
    \frac{e^{-G^2/(4\alpha^2)}}{G^2}\,|S(\mathbf{G})|^2.
\end{equation}
This form incurs the truncation error
\begin{equation}
\begin{split}
    \Delta E
    &=
    E_{\mathrm{Ewald}} - E_{\mathrm{Ewald}}^{R_c,G_c}
    \\
    &=
    \underbrace{\left(E_{\mathrm{real}}^{\infty}-E_{\mathrm{real}}^{R_c}\right)}_{
    \Delta E_{\mathrm{real\text{-}tail}}}
    +
    \underbrace{\left(E_{\mathrm{recip}}^{\infty}-E_{\mathrm{recip}}^{G_c}\right)}_{
    \Delta E_{\mathrm{recip\text{-}tail}}}.
\end{split}
\label{eq:ewald_tail_decomp}
\end{equation}
Increasing \(\alpha\) speeds up the real-space decay and suppresses \(\Delta E_{\mathrm{real\text{-}tail}}\), but raises the reciprocal-space cutoff \(G_c\) needed for a fixed reciprocal-space error \(\Delta E_{\mathrm{recip\text{-}tail}}\).

\begin{figure*}[t]
    \centering
    \includegraphics[width=\textwidth]{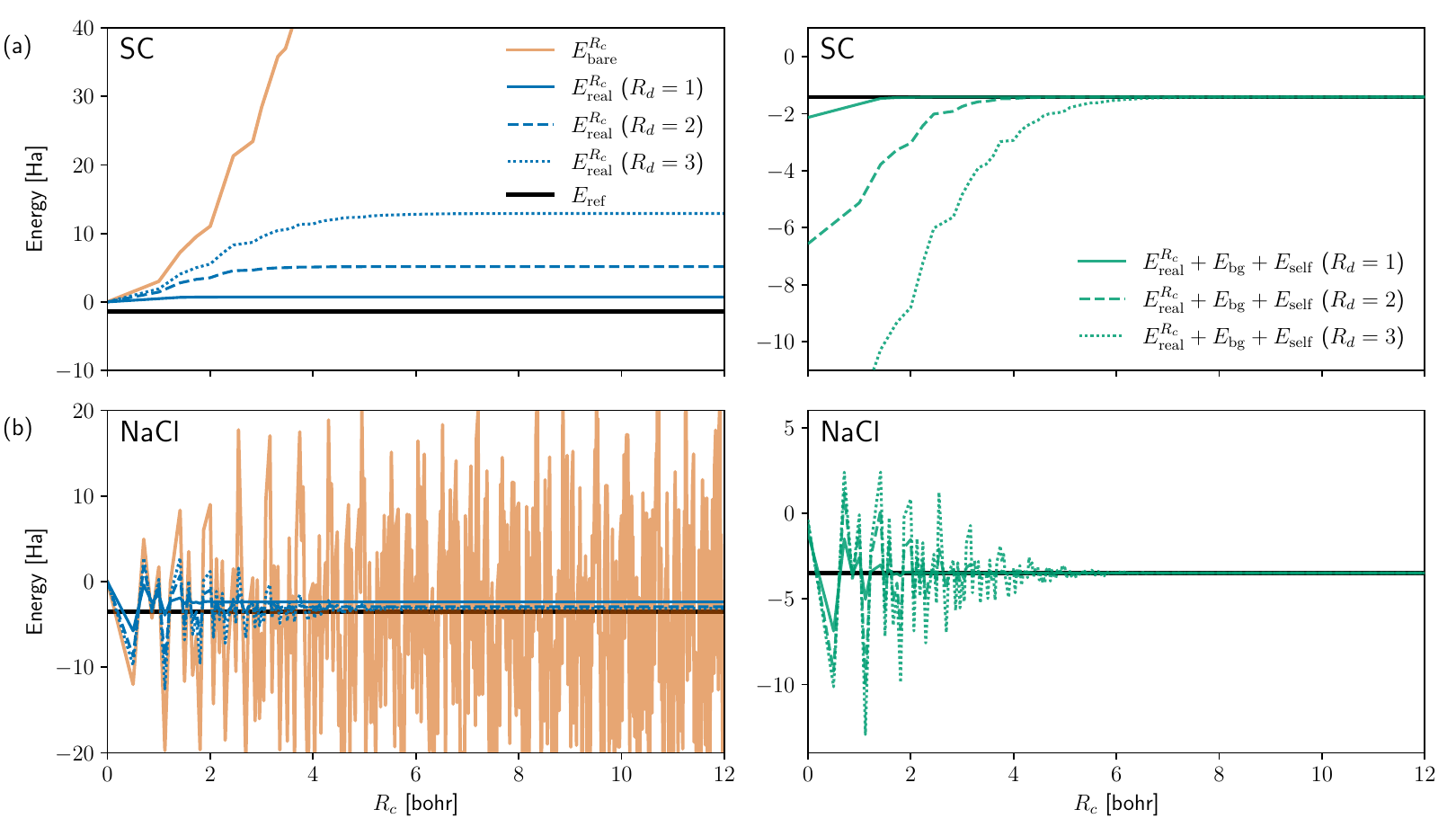}
    \caption{
        \textbf{Convergence behavior of terms in the localized Ewald sum.}
        Results are shown for
        (a) simple cubic (SC) structure with \(a=1\)~bohr and \(q=+1\), and
        (b) NaCl structure with \(a=1\)~bohr and \(q_1=+1, q_2=-1\).
        The solid black line in each panel marks the reference energy \(E_{\mathrm{ref}}\) from a converged Ewald evaluation for SC and from the Madelung reference of Ref.~\protect\citep{mamode2017computation} for NaCl.
        The left column shows the bare and damped pair sums, \(E_{\mathrm{bare}}^{R_c}\) and \(E_{\mathrm{real}}^{R_c}\), where the former diverges and the latter converges but to the wrong result.
        The right column shows the damped real-space sum together with the self and background corrections, \(E_{\mathrm{real}}^{R_c} + E_{\mathrm{bg}}+E_{\mathrm{self}}\), which achieves the correct result for large \(R_c\).
    }
    \label{fig:local_ewald_terms}
\end{figure*}

\section{Localized Ewald Approximation}

To facilitate later analysis, we consider a simple real-space approximation to the Ewald sum, defining the localized Ewald energy,
\begin{equation}
    E_{\mathrm{LE}}=
    E_{\mathrm{real}}^{R_c} + E_{\mathrm{self}} + E_{\mathrm{bg}},
\end{equation}
which approaches the exact electrostatic energy for sufficiently large \(R_d\) and \(R_c\).
Since \(E_{\mathrm{LE}}\) omits the reciprocal-space contribution entirely, its error relative to \(E_{\mathrm{Ewald}}\) contains the full reciprocal-space contribution \(E_{\mathrm{recip}}^{\infty}\) rather than just the reciprocal tail \(\Delta E_{\mathrm{recip\text{-}tail}}\),
\begin{equation}
    \Delta E_{\mathrm{LE}}
    =
    E_{\mathrm{Ewald}} - E_{\mathrm{LE}}
    =
    \Delta E_{\mathrm{real\text{-}tail}}
    +E_{\mathrm{recip}}^{\infty}.
\label{eq:local_error}
\end{equation}

To characterize the convergence behavior of the localized Ewald approximation, we contrast the ``bare'' pair sum
\begin{equation}
    E_{\mathrm{bare}}^{R_c}=
    \frac12\sum_{i=1}^{N_\text{cell}}\sum_{\substack{j\ne i\\ r_{ij}\le R_c}}\frac{q_i q_j}{r_{ij}}
\end{equation}
with the ``damped'' real-space sum \(E_{\mathrm{real}}^{R_c}\) of Eq.~\eqref{eq:ewald_real_rc}, which replaces the bare \(1/r\) interaction with the damped kernel \(\phi_{R_d}\).
The convergence behavior of both sums, together with the self and background terms, is shown in Figure~\ref{fig:local_ewald_terms} for the simple cubic and NaCl structures.
Except for NaCl, which carries charges \(\pm 1\), every test system in this work carries only positive charges and is held neutral by the uniform background of \(E_{\mathrm{bg}}\).
The bare pair sum diverges with \(R_c\) for the simple cubic structure because the slowly decaying \(1/r\) potential causes the sum over like charges to grow without bound,
while it oscillates without convergence for the NaCl structure because of the alternating signs of the charges.
In fact, electrostatic sums for the three-dimensional NaCl structure, when evaluated via expanding spheres, are known to diverge~\citep{borwein1985convergence}.
In general, such summations are conditionally convergent, depending on the details of the summation order.
Wolf~\citep{wolf1999exact} and Pickard~\citep{PhysRevMaterials.2.013806} reported similar observations before introducing their local charge-neutralization schemes to make the bare pair sum converge.
Damping was introduced later to accelerate convergence.

In the localized Ewald approximation, there are no terms that induce local charge neutrality.
The self and background terms only provide a global shift.
Figure~\ref{fig:local_ewald_terms} shows that damping is sufficient for the real-space sum to converge as \(R_c\) increases (left column). Convergence alone does not reproduce the Ewald energy, however.
The self and background terms provide the shift that brings the converged value into agreement with the reference energy \(E_{\mathrm{ref}}\) (right column),
up to the omitted reciprocal contribution \(E_{\mathrm{recip}}^{\infty}\) of Eq.~\eqref{eq:local_error}.
If damping is sufficient for convergence, what roles do the local charge-neutralization schemes of Wolf and Pickard play?
We demonstrate that they can be viewed as approximations to the finite-cutoff truncation error of the real-space electrostatic sum.

\section{Wolf Sum}
\label{sec:wolf_sum}

For a neutral cell (\(Q=0\)), the damped Wolf sum~\citep{wolf1999exact} is
\begin{equation}
    E_{\mathrm{Wolf}}(R_c)
    =
    E_{\mathrm{real}}^{R_c} + E_{\mathrm{self}} + E_{\mathrm{shell}}(R_c),
\end{equation}
where the shell term
\begin{equation}
    E_{\mathrm{shell}}(R_c)
    =
    -\frac{1}{2}\sum_{i=1}^{N_\text{cell}} q_i\, Q_i(R_c)\,\phi_{R_d}(R_c)
\label{eq:wolf_shell}
\end{equation}
places a compensating charge \(-Q_i(R_c)\) on the cutoff sphere of each atom \(i\), and
\begin{equation}
    Q_i(r)
    =
    \sum_{\substack{j\\ r_{ij}\le r}} q_j
    \quad \text{(including \(j=i\))}.
\end{equation}
We now relate the shell term to the real-space truncation error \(\Delta E_{\mathrm{real\text{-}tail}}\) of Eq.~\eqref{eq:ewald_tail_decomp}.
Since \(\phi_{R_d}(r)\to 0\) as \(r\to\infty\), each pair term can be written as \(\phi_{R_d}(r_{ij})=-\int_{r_{ij}}^{\infty}\phi'_{R_d}(r)\,\mathrm{d}r\).
Summing over the neighbors beyond \(R_c\) and exchanging the sum with the integral gives
\begin{equation}
\begin{split}
    \sum_{\substack{j\\ r_{ij}>R_c}} q_j\,\phi_{R_d}(r_{ij})
    &=
    -\int_{R_c}^{\infty}
    \Biggl[\,\sum_{\substack{j\\ R_c<r_{ij}\le r}} q_j\Biggr]
    \phi'_{R_d}(r)\,\mathrm{d}r
    \\
    &=
    -\int_{R_c}^{\infty}\left[Q_i(r)-Q_i(R_c)\right]\phi'_{R_d}(r)\,\mathrm{d}r
    \\
    &=
    -Q_i(R_c)\,\phi_{R_d}(R_c)
    \\
    &\quad
    -\int_{R_c}^{\infty}Q_i(r)\,\phi'_{R_d}(r)\,\mathrm{d}r,
\end{split}
\label{eq:wolf_tail_identity}
\end{equation}
where the exchange is allowed because the damped sum converges absolutely, and the integral in the last line converges because \(|Q_i(r)|\) grows at most as \(r^3\).
Weighting Eq.~\eqref{eq:wolf_tail_identity} by \(q_i/2\) and summing over \(i\) then gives
\begin{equation}
    \Delta E_{\mathrm{real\text{-}tail}}
    =
    E_{\mathrm{shell}}(R_c)
    -\frac12\sum_{i=1}^{N_\text{cell}} q_i\int_{R_c}^{\infty}Q_i(r)\,\phi'_{R_d}(r)\,\mathrm{d}r.
\label{eq:wolf_trunc}
\end{equation}
The shell term, introduced by Wolf for charge neutralization, is thus the boundary term of the real-space tail, with a remainder that averages the oscillating enclosed charge beyond \(R_c\) with weight \(-\phi'_{R_d}(r)>0\).

\section{Pickard Sum}
\label{sec:pickard_sum}

We now show that Pickard's corrections also approximate \(\Delta E_{\mathrm{real\text{-}tail}}\).
The total energy in Pickard's method is
\begin{equation}
    E_{\mathrm{Pickard}}
    =
    E_{\mathrm{real}}^{R_c}
    +E_{\mathrm{sphere}}
    +E_{\mathrm{damp}}
    +E_{\mathrm{self}}.
\end{equation}
Pickard introduced the terms
\begin{equation}
    E_{\mathrm{sphere}} = \sum_{i=1}^{N_\text{cell}} E_{i,\mathrm{sphere}}
    \quad \text{and} \quad
    E_{\mathrm{damp}}=\sum_{i=1}^{N_\text{cell}} E_{i,\mathrm{damp}},
\end{equation}
through a local neutralizing-sphere construction, where the per-charge terms
\begin{equation}
 E_{i,\mathrm{sphere}}
=
- \pi \rho q_i R_{a,i}^2
\end{equation}
and
\begin{equation}
\begin{split}
    E_{i,\mathrm{damp}}
    &=
    \pi \rho q_i\!\left(R_{a,i}^2-\frac{R_d^2}{2}\right)\!
    \operatorname{erf}(R_{a,i}/R_d)\\
    &\quad
    +\sqrt{\pi}\,\rho q_i R_{a,i}R_d\,e^{-R_{a,i}^2/R_d^2}
\end{split}
\end{equation}
are contributions from the uniform charge distribution within this sphere, decomposed into the bare Coulomb part and the damping correction, respectively.
A simple choice for the compensation sphere radius \(R_a\) is to set it equal to the real-space cutoff \(R_c\).
Instead, Pickard's original formulation sizes the sphere adaptively based on the local enclosed charge to impose local charge neutrality,
\begin{equation}
    \frac{4\pi}{3}R_{a,i}^3\rho
    =
    \sum_{\substack{j\\ r_{ij}\le R_c}} q_j \quad \text{(including \(j=i\))}
    ,
\label{eq:Ra_adaptive}
\end{equation}
where 
\begin{equation}
    \rho = \frac{1}{\Omega} \sum_{i=1}^{N_\text{cell}} q_i = \frac{Q}{\Omega}
\end{equation}
is the average charge density.
With charges of both signs present, the positive and negative charges are neutralized separately.
Each group enters the expressions above with its own partial density \(\rho_\pm=\Omega^{-1}\sum_{i\in\pm}q_i\) and adaptive radius \(R^{\pm}_{a,i}\), and the compensation terms are summed over both.
For simplicity, the analysis below assumes charges of a single sign.

Compared to the localized Ewald sum, the compensation terms \(E_{\mathrm{sphere}}\) and \(E_{\mathrm{damp}}\) together replace the global neutralizing background \(E_{\mathrm{bg}}\) of Eq.~\eqref{eq:ewald_bg} with localized sphere-based corrections.
To relate these terms, we note that the background term represents the interaction of each charge \(q_i\) with the compensating uniform charge density \(-\rho\) via the damped kernel \(\phi_{R_d}\), integrated over all space.
The per-charge form is
\begin{equation}
    E_{i,\mathrm{bg}}
    =
    \frac{q_i}{2}\int (-\rho) \phi_{R_d}(r) \,\mathrm{d}^3r
    = -\frac{\pi}{2}\,q_i\rho R_d^2.
\end{equation}
Pickard's correction terms arise from the same interaction, restricted to a sphere of radius \(R_{a,i}\),
\begin{equation}
    E_{i,\mathrm{sphere}} + E_{i,\mathrm{damp}}
    =
    \frac{q_i}{2}\!\int_{|\mathbf{r}|<R_{a,i}}\! (-\rho) \phi_{R_d}(r) \,\mathrm{d}^3r.
\end{equation}

\begin{figure*}[t]
    \centering
    \includegraphics[width=\textwidth]{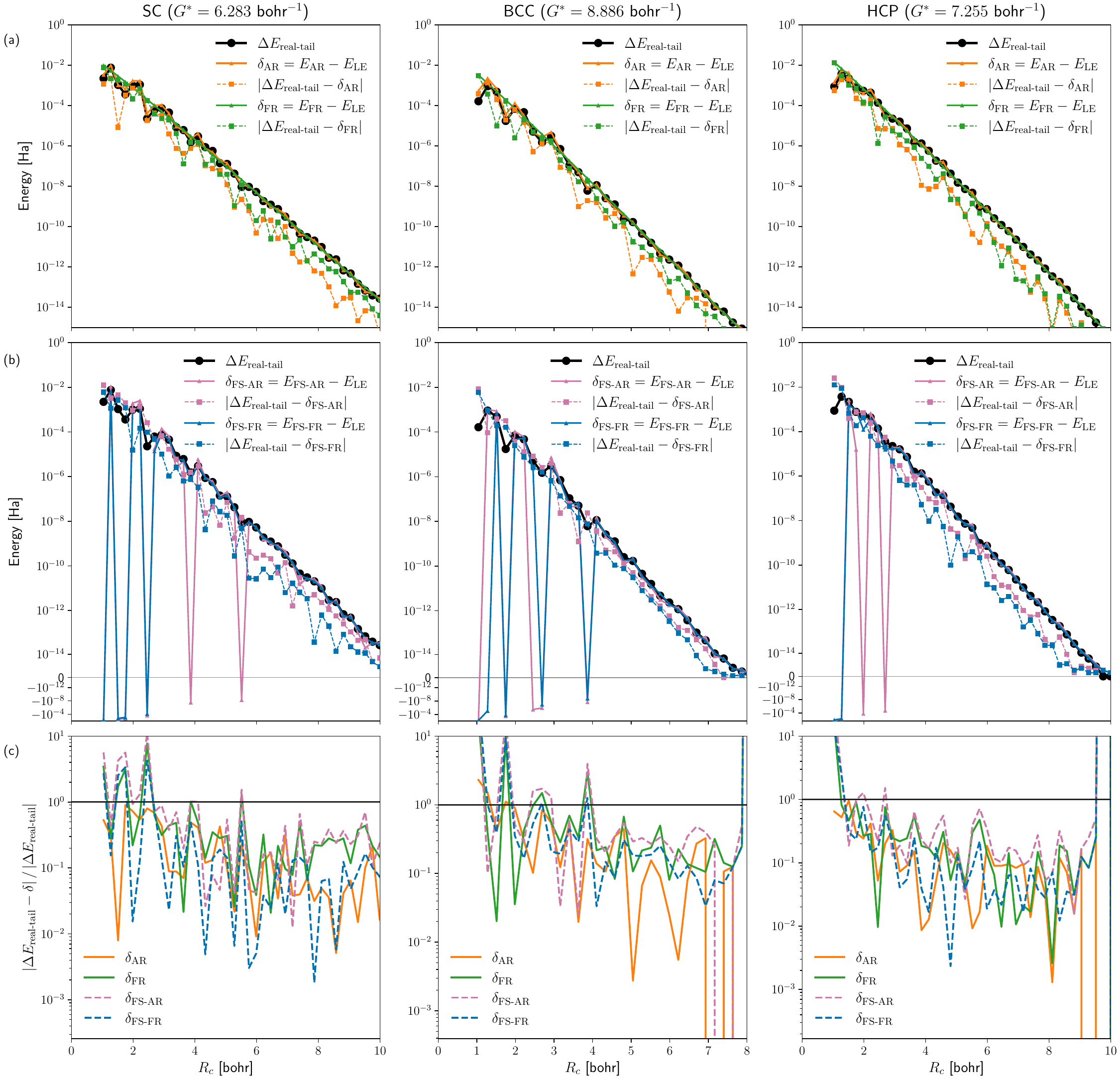}
    \caption{\textbf{Truncation-error compensation behavior of Pickard variants.}
    Results are shown for simple cubic (SC), body-centered cubic (BCC), and ideal hexagonal close-packed (HCP) structures (\(a=1\)~bohr, \(q=+1\) per atom), with \(R_d\) scaled with \(R_c\) using the parameter scaling relation of Eq.~\eqref{eq:optimal_scaling}.
    Let \(\delta_X = E_X - E_{\mathrm{LE}}\) be the energy difference between Pickard variant \(X\) and localized Ewald.
    Row (a) compares \(\delta_X\) with \(\Delta E_{\mathrm{real\text{-}tail}}\) for adaptive-radii Pickard (AR-Pickard) and fixed-radii Pickard (FR-Pickard), together with the residuals \(|\Delta E_{\mathrm{real\text{-}tail}}-\delta_X|\).
    Row (b) shows the same for force-shifted AR-Pickard (FS-AR-Pickard) and force-shifted FR-Pickard (FS-FR-Pickard).
    Row (c) shows the ratio \(|\Delta E_{\mathrm{real\text{-}tail}}-\delta_X|/|\Delta E_{\mathrm{real\text{-}tail}}|\) for all four variants.
    }
    \label{fig:pickard_error_analysis}
\end{figure*}

The difference between \(E_{i,\mathrm{sphere}} + E_{i,\mathrm{damp}}\) and \(E_{i,\mathrm{bg}}\) is therefore an integral over the complementary outer region,
\begin{equation}
    (E_{i,\mathrm{sphere}} + E_{i,\mathrm{damp}}) - E_{i,\mathrm{bg}}
    = \frac{q_i}{2}\int_{|\mathbf{r}|>R_{a,i}} \rho \phi_{R_d}(r) \,\mathrm{d}^3r.
\label{eq:pickard_minus_bg}
\end{equation}
This outer-tail form points directly to the real-space truncation error of the Ewald sum \(\Delta E_{\mathrm{real\text{-}tail}}=E_{\mathrm{real}}^{\infty}-E_{\mathrm{real}}^{R_c}\) from Eq.~\eqref{eq:ewald_tail_decomp}.
The per-charge contribution to the real-space tail is
\begin{equation}
    \Delta E_{i,\mathrm{tail}}
    =
    \frac{q_i}{2} \sum_{\substack{j\\ r_{ij}>R_c}} q_j\,\phi_{R_d}(r_{ij}),
\end{equation}
where the index set is taken to include \(j=i\), since the self term does not contribute for \(R_c>0\).
Using an equivalent charge-density representation,
\begin{equation}
    \Delta E_{i,\mathrm{tail}}
    =
    \frac{q_i}{2} \int_{|\mathbf{r}|>R_c} \rho_i(\mathbf{r})\,\phi_{R_d}(r)\,\mathrm{d}^3r,
\end{equation}
where
\begin{equation}
    \rho_i(\mathbf{r}) = \sum_j q_j \delta\!\left(\mathbf{r}-(\mathbf{r}_j-\mathbf{r}_i)\right),
\end{equation}
we can decompose \(\rho_i\) into its uniform and oscillatory parts,
\begin{equation}
    \rho_i(\mathbf{r})=
    \rho
    +
    \sum_{\mathbf{G}\neq \mathbf{0}}\rho_{i,\mathbf{G}}e^{i\mathbf{G}\cdot\mathbf{r}},
\end{equation}
where \(\rho=Q/\Omega\) is the average charge density defined earlier.
Introducing a family of integrals for the corresponding uniform real-space tail contribution,
\begin{equation}
    \Delta E_{i,\mathrm{tail}}^{(G=0)}(R)
    \equiv
    \frac{q_i}{2}\!\int_{|\mathbf{r}|>R}\! \rho \phi_{R_d}(r)\,\mathrm{d}^3r,
\end{equation}
we split the truncation error as
\begin{equation}
    \Delta E_{i,\mathrm{tail}}
    =
    \Delta E_{i,\mathrm{tail}}^{(G=0)}(R_c) + \Delta E_{i,\mathrm{tail}}^{(G\neq 0)}.
\end{equation}
From Eq.~\eqref{eq:pickard_minus_bg}, the Pickard correction supplies precisely \(\Delta E_{i,\mathrm{tail}}^{(G=0)}(R_{a,i})\).
The error of the Pickard sum relative to the full Ewald sum is therefore
\begin{equation}
\begin{split}
    \Delta E_{\mathrm{Pickard}}
    =
    &\sum_{i=1}^{N_\text{cell}}\!\left[\Delta E_{i,\mathrm{tail}}^{(G=0)}(R_c)-\Delta E_{i,\mathrm{tail}}^{(G=0)}(R_{a,i})\right]  \\
    &+\sum_{i=1}^{N_\text{cell}}\Delta E_{i,\mathrm{tail}}^{(G\neq 0)} +E_{\mathrm{recip}}^{\infty},
\end{split}
\label{eq:pickard_error}
\end{equation}
which includes the finite-radius mismatch of the uniform tail contribution, the residual oscillatory real-space tail beyond \(R_c\), and the omitted reciprocal contribution \(E_{\mathrm{recip}}^{\infty}\).
Relative to localized Ewald, the Pickard sum additionally compensates the uniform part of the real-space truncation error.

This truncation-error compensation behavior is illustrated in Figure~\hyperref[fig:pickard_error_analysis]{\ref*{fig:pickard_error_analysis}(a)} for simple cubic (SC), body-centered cubic (BCC), and ideal hexagonal close-packed (HCP) structures along the optimal parameter path (described in Sec.~\ref{sec:error_scaling}).
In all three systems, the difference between the adaptive-radii Pickard (AR-Pickard) and localized Ewald energies, \(\delta_{\mathrm{AR}} = E_{\mathrm{AR}} - E_{\mathrm{LE}}\), tracks the real-space tail error \(\Delta E_{\mathrm{real\text{-}tail}}\) closely with the same sign and magnitude.
The residual difference \(|\Delta E_{\mathrm{real\text{-}tail}} - \delta_{\mathrm{AR}}|\) is typically an order of magnitude smaller than \(|\Delta E_{\mathrm{real\text{-}tail}}|\), indicating that Pickard's correction captures the leading part of the real-space truncation error.

\begin{figure*}[t]
    \centering
    \includegraphics[width=\textwidth]{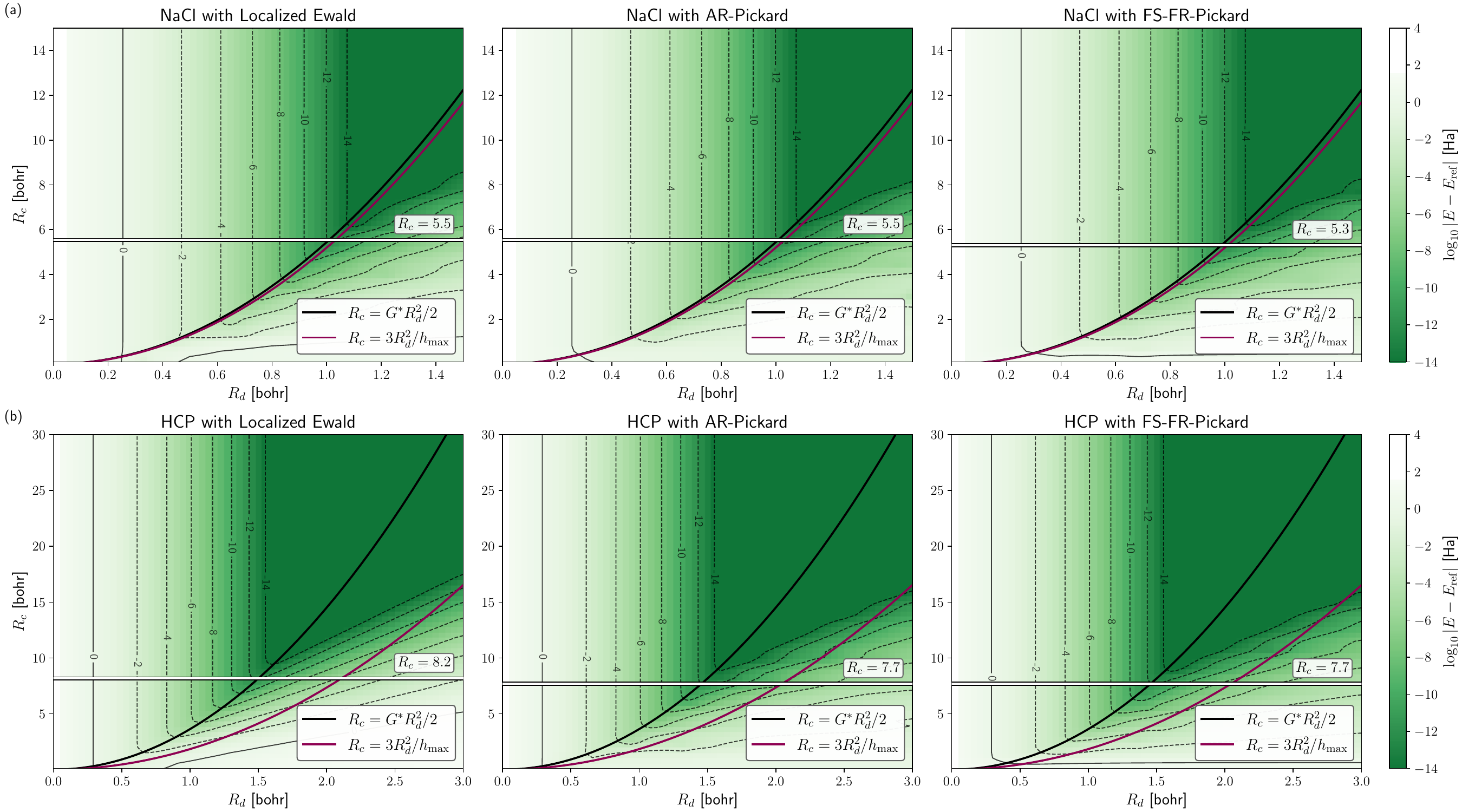}
    \caption{\textbf{Convergence behavior in parameter space for localized Ewald and Pickard variants.}
    Error maps of \(\log_{10}|E-E_{\mathrm{ref}}|\) in the \((R_d,R_c)\) plane are shown for localized Ewald, adaptive-radii Pickard (AR-Pickard), and force-shifted fixed-radii Pickard (FS-FR-Pickard), for
    (a) NaCl structure (\(a=1\)~bohr, \(q_1=+1, q_2=-1\)) with reference energy \(E_{\mathrm{ref}}\) from Ref.~\protect\citep{mamode2017computation}, and
    (b) ideal hexagonal close-packed (HCP) structure (\(a=1\)~bohr, \(q_1=q_2=+1\)) with reference energy \(E_{\mathrm{ref}}\) from Ref.~\protect\citep{PhysRevE.107.065302}.
    The derived scaling relation \(R_c=G^{*}R_d^2/2\) (black) and Pickard's empirical relation \(R_c=3R_d^2/h_{\max}\) (magenta) are shown.
    Guide lines mark the minimum \(R_c\) at which the error first reaches \(10^{-12}\)~Ha.
    }
    \label{fig:nacl_hcp_scaling}
\end{figure*}

\section{Modified Pickard Sum}

The Pickard sum admits two independently applicable modifications.
The first, motivated by the error analysis of Sec.~\ref{sec:pickard_sum}, is a simplification that fixes the compensation sphere radius to \(R_{a,i}=R_c\) for all atoms.
The second replaces the damped Coulomb kernel \(\phi_{R_d}\) with a force-shifted one \(\tilde{\phi}_{R_d}\) to eliminate energy and force discontinuities at the cutoff that may cause unphysical behavior during dynamical simulations.
We refer to the four resulting combinations as adaptive-radii Pickard (AR-Pickard), fixed-radii Pickard (FR-Pickard), force-shifted AR-Pickard (FS-AR-Pickard), and force-shifted FR-Pickard (FS-FR-Pickard).
AR-Pickard is Pickard's original formulation~\citep{PhysRevMaterials.2.013806}, and the remaining three are defined in this section.

\textbf{Fixed-Radii Simplification}.
Setting \(R_{a,i}=R_c\) for all atoms exactly balances the \(G=0\) mismatch term of Eq.~\eqref{eq:pickard_error}, leaving only the \(G\neq 0\) real-space tail and \(E_{\mathrm{recip}}^{\infty}\) as sources of error.
The correction potential \((E_{i,\mathrm{sphere}}+E_{i,\mathrm{damp}})/q_i\) becomes a per-system constant independent of the local charge environment, requiring no per-atom neighbor accumulation and making FR-Pickard more computationally efficient than AR-Pickard.
These two variants are compared in Figure~\hyperref[fig:pickard_error_analysis]{\ref*{fig:pickard_error_analysis}(a)}, where both track \(\Delta E_{\mathrm{real\text{-}tail}}\) closely in sign and magnitude across all three systems.
The two variants show no significant systematic difference in truncation-error compensation quality.

\textbf{Imposing Force Continuity}.
In the real-space pair sum, the damped kernel [Eq.~\eqref{eq:phi_def}] is truncated at \(R_c\).
The abrupt truncation creates a jump in both the pair energy kernel and its radial derivative at \(R_c\), so both the pair energy and force are discontinuous when a pair crosses the cutoff, which could be a source of unphysical behavior in dynamical simulations.
To remove both discontinuities, we replace \(\phi_{R_d}\) inside the cutoff by a force-shifted kernel~\citep{fennell2006ewald} obtained by subtracting the first-order Taylor expansion of \(\phi_{R_d}(r)\) about \(r=R_c\),
\begin{equation}
    \tilde\phi_{R_d}(r)
    = \phi_{R_d}(r)
    - \phi_{R_d}(R_c)
    - (r-R_c)\,\phi'_{R_d}(R_c),
\label{eq:fs_kernel}
\end{equation}
where
\begin{equation}
    \phi'_{R_d}(R_c)
    = -\frac{\phi_{R_d}(R_c)}{R_c}
    -\frac{2\,e^{-(R_c/R_d)^2}}{\sqrt{\pi}\,R_d\,R_c}.
\end{equation}
By construction, \(\tilde\phi_{R_d}(R_c)=0\) and \(\tilde\phi'_{R_d}(R_c)=0\), so both the pair energy and force are continuous at \(R_c\).
Replacing \(\phi_{R_d}\) by \(\tilde\phi_{R_d}\) in the pair sum of Pickard's method [Eq.~\eqref{eq:ewald_real_rc}] adds the term
\begin{equation}
\begin{split}
    E_{\mathrm{shift}}
    &= - \frac{1}{2}\sum_{i=1}^{N_\text{cell}} q_i \sum_{\substack{j\ne i\\ r_{ij}\le R_c}} q_j
    \\
    &\quad\times
    \left[ \phi_{R_d}(R_c) + (r_{ij} - R_c)\,\phi'_{R_d}(R_c) \right].
\end{split}
\end{equation}
Direct inclusion of this new term, however, degrades the compensation of the uniform tail contribution achieved by Pickard's method (see the Supplementary Information for an example).
To counteract this effect, we introduce an additional term derived from a uniform-density approximation of \(E_{\mathrm{shift}}\).
Defining
\begin{equation}
    f(r)=\phi_{R_d}(R_c)+(r-R_c)\phi'_{R_d}(R_c),
\end{equation}
and using
\begin{equation}
    \sum_{\substack{j\ne i\\r_{ij}\le R_c}} q_j f(r_{ij})
    =
    \sum_{\substack{j\\r_{ij}\le R_c}} q_j f(r_{ij})-q_i f(0),
\end{equation}
we approximate the sum of neighboring charges by a uniform-density integral, yielding the correction term
\begin{align}
    E_{\mathrm{shift\text{-}corr}}
    &= \frac{1}{2}\sum_{i=1}^{N_\text{cell}} q_i
    \left[
    \rho\!\int_0^{R_c}\!f(r)\,4\pi r^2\,\mathrm{d}r
    -q_i f(0)
    \right]
    \nonumber\\
    &= \frac{\rho Q}{2}\cdot \frac{\pi R_c^3}{3}\bigl[4\,\phi_{R_d}(R_c) - R_c\,\phi'_{R_d}(R_c)\bigr]
    \nonumber\\
    &\quad
    -\frac{1}{2}\sum_{i=1}^{N_\text{cell}} q_i^2\bigl[\phi_{R_d}(R_c)-R_c\phi'_{R_d}(R_c)\bigr].
\end{align}
We thus introduce the force-shifted Pickard sum
\begin{equation}
    E_{\mathrm{FS\text{-}Pickard}}
    =
    E_{\mathrm{Pickard}} + E_{\mathrm{shift}} + E_{\mathrm{shift\text{-}corr}}.
\end{equation}
For \(Q=0\), the FS-FR variant reduces exactly to the damped shifted-force method~\citep{fennell2006ewald}, including the self-interaction term of its LAMMPS implementation~\citep{thompson2022lammps}.
For FS-FR, the remaining terms are independent of atomic positions, so the total energy and forces are continuous as well.
For FS-AR, the adaptive radius \(R_{a,i}\) is set by the enclosed charge in Eq.~\eqref{eq:Ra_adaptive}, which changes abruptly when a neighbor crosses \(R_c\), so the total energy remains discontinuous.
A continuous attenuation of the enclosed charge near \(R_c\) would remove this jump, but it also alters the local charge balance of Eq.~\eqref{eq:Ra_adaptive}.
This is beyond the scope of the present work.

Figure~\hyperref[fig:pickard_error_analysis]{\ref*{fig:pickard_error_analysis}(b)} shows the FS variants for the same three systems.
Although more oscillatory in sign at smaller \(R_c\) because of the additional terms, \(\delta_{\mathrm{FS\text{-}AR/FR}} = E_{\mathrm{FS\text{-}AR/FR}}-E_{\mathrm{LE}}\) still follows \(\Delta E_{\mathrm{real\text{-}tail}}\) closely in both sign and magnitude at moderate \(R_c\) values, and \(|\Delta E_{\mathrm{real\text{-}tail}}-\delta_{\mathrm{FS\text{-}AR/FR}}|\) remains roughly an order of magnitude smaller than \(|\Delta E_{\mathrm{real\text{-}tail}}|\) across all three systems.
Together, \(E_{\mathrm{shift}}\) and \(E_{\mathrm{shift\text{-}corr}}\) thus preserve the truncation-error compensation behavior of the unshifted variants.

\section{Optimal Parameter Scaling}
\label{sec:error_scaling}

Damped real-space methods, including the localized Ewald, damped Wolf, and Pickard methods, are controlled by the same two parameters: the real-space cutoff \(R_c\) and the screening distance \(R_d=1/\alpha\).
For practical calculations, one seeks a relation between \(R_c\) and \(R_d\) that reaches a target accuracy at minimal cost.
Practitioners of Wolf's method have typically selected them through empirical parameter sweeps and heuristics tailored to specific systems~\citep{demontis2001application,zahn2002enhancement,hansen2012simplistic}.
Pickard identified an empirical relation for the optimal path to convergence in the \((R_d,R_c)\) parameter space for his method, given by
\begin{equation}
R_c\approx 3\,\frac{R_d^2}{h_{\max}},
\end{equation}
where \(h_{\max}\) is the largest perpendicular distance between the faces of the primitive cell or a chosen characteristic system length scale~\citep{PhysRevMaterials.2.013806}.
Here, we derive a more general nonempirical form that applies to the class of damped real-space sums by balancing the leading asymptotic scales of the omitted real-space tail and reciprocal-space contributions.

These real-space methods omit the same reciprocal contribution \(E_{\mathrm{recip}}^{\infty}\).
Grouping by reciprocal shells (\(G=|\mathbf{G}|\)) gives
\begin{equation}
    E_{\mathrm{recip}}^{\infty}
    =\sum_{\text{shells }G}
    C_{\mathrm{shell}}(G)\,e^{-(G/(2\alpha))^2},
\end{equation}
with the reciprocal-shell electrostatic coefficient
\begin{equation}
    C_{\mathrm{shell}}(G)\equiv
    \frac{2\pi}{\Omega}\sum_{\mathbf{G}\in \mathrm{shell}(G)}
    \frac{|S(\mathbf{G})|^2}{G^2}.
\end{equation}
As \(\alpha\) decreases, shells at larger \(G\) are exponentially suppressed by \(\exp[-(G/(2\alpha))^2]\), so the omitted reciprocal contribution \(E_{\mathrm{recip}}^{\infty}\) is asymptotically dominated by the smallest shell with nonnegligible \(C_{\mathrm{shell}}\), denoted \(G^{*}\).
Shell locations are set by reciprocal-lattice geometry, while shell activity is set by the charge-weighted structure factor \(S(\mathbf{G})\), so \(G^{*}\) depends on both lattice geometry and charge arrangement.
The leading reciprocal-space contribution is therefore
\begin{equation}
    E_{\mathrm{recip}}^{\infty}
    \sim
    C_{\mathrm{shell}}(G^{*})\exp\!\left(-\left(\frac{G^{*}}{2\alpha}\right)^{\!2}\right).
\end{equation}

For the real-space tail, we consider the \(R_c\to\infty\) limit and use \(\operatorname{erfc}(x) \approx e^{-x^2}/(\sqrt{\pi}x)\) for \(x\gg 1\) to obtain
\begin{equation}
    \Delta E_{\mathrm{real\text{-}tail}}
    \approx
    e^{-(\alpha R_c)^2}
    \underbrace{\left(
    \sum_{\substack{i,j\\r_{ij}>R_c}}
    \frac{q_i q_j\,e^{-\alpha^2(r_{ij}^2-R_c^2)}}{2\sqrt{\pi}\,\alpha\,r_{ij}^2}
    \right)}_{A_0}.
\end{equation}
The largest contributions to the real-space tail come from pairs near \(R_c\), where \(r_{ij}^2-R_c^2\ll R_c^2\).
We therefore expect the \(\alpha\)-dependence within \(A_0\) to be much weaker than that of \(e^{-(\alpha R_c)^2}\), which sets the leading exponential scale.

The real-space truncation error of the damped Wolf and Pickard-family variants shares the same leading exponential scale \(e^{-(\alpha R_c)^2}\), with method-dependent prefactors \(A\).
For the Wolf sum, the shell term [Eq.~\eqref{eq:wolf_shell}] is proportional to \(\phi_{R_d}(R_c)=\operatorname{erfc}(\alpha R_c)/R_c\), which carries the same \(e^{-(\alpha R_c)^2}\) scale as \(\Delta E_{\mathrm{real\text{-}tail}}\) for large \(R_c\), so the remainder in Eq.~\eqref{eq:wolf_trunc} does too.
As shown in Sec.~\ref{sec:pickard_sum}, the Pickard correction [Eq.~\eqref{eq:pickard_minus_bg}] is an integral of \(\phi_{R_d}(r)\) beyond \(R_{a,i}\approx R_c\) and carries the same scale.
Furthermore, its partial compensation of \(\Delta E_{\mathrm{real\text{-}tail}}\) is expected to reduce \(A\) relative to \(A_0\).

\begin{figure*}[t]
    \centering
    \includegraphics[width=\textwidth]{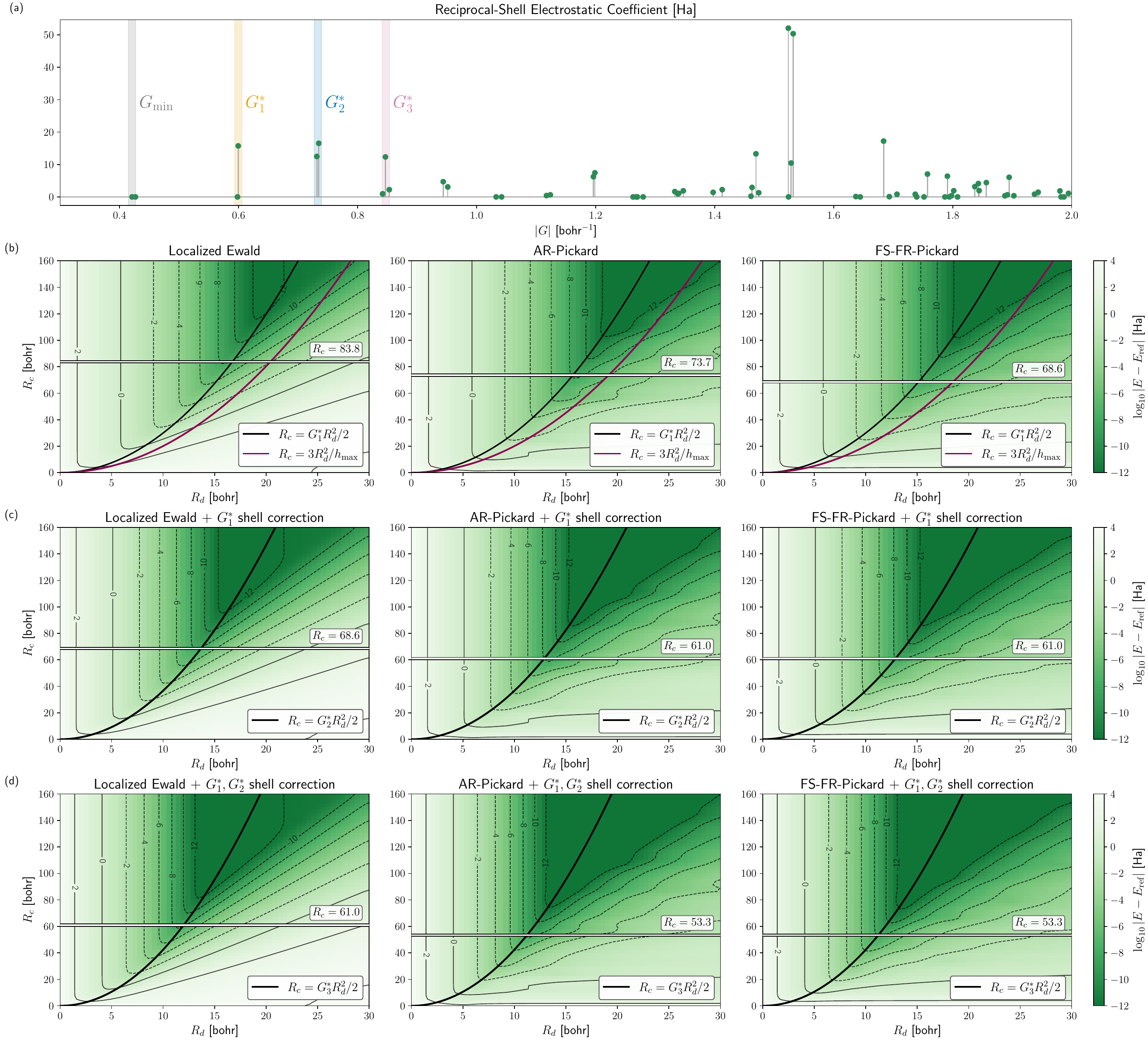}
    \caption{\textbf{Convergence behavior of localized Ewald and Pickard variants for Al\(_2\)SiO\(_5\).}
    (a) Reciprocal-shell electrostatic coefficient against \(G\), showing extinction at \(G_{\min}\) and active shells \(G_1^{*}\), \(G_2^{*}\), and \(G_3^{*}\), where \(G_2^{*}\) and \(G_3^{*}\) are coefficient-weighted means of two and three nearly coincident shells, respectively.
    Error maps of \(\log_{10}|E-E_{\mathrm{ref}}|\) in the \((R_d,R_c)\) plane for localized Ewald, adaptive-radii Pickard (AR-Pickard), and force-shifted fixed-radii Pickard (FS-FR-Pickard), with converged Ewald reference energy \(E_{\mathrm{ref}}=-244.055008509077\)~Ha, for
    (b) the uncorrected real-space methods,
    (c) \(G_1^{*}\)-shell-corrected methods, and
    (d) \(G_1^{*}\!+\!G_2^{*}\)-shell-corrected methods.
    The derived scaling relation \(R_c=(G_i^{*}/2)R_d^2\) (black) and Pickard's empirical relation \(R_c=3R_d^2/h_{\max}\) (magenta, \(h_{\max}=14.93\)~bohr) are shown.
    Guide lines mark the minimum \(R_c\) at which the error first reaches \(10^{-8}\)~Ha.
    The valence charges used are \(q_\mathrm{Al}=+3\), \(q_\mathrm{Si}=+4\), and \(q_\mathrm{O}=+6\).
    }
    \label{fig:al2sio5_analysis}
\end{figure*}

Combining this real-space error contribution with the omitted reciprocal contribution, we model the total error as
\begin{equation}
    \varepsilon(\alpha)
    =
    A e^{-(\alpha R_c)^2}
    +
    B e^{-(G^{*}/(2\alpha))^2},
\label{eq:total_error}
\end{equation}
where \(B=C_{\mathrm{shell}}(G^{*})\).
To obtain the leading optimal scaling, we treat \(A\) as slowly varying with \(\alpha\) compared with the exponential factors.
In the asymptotic regime \(\lambda = R_c G^{*}/2 \gg 1\), the optimal scaling path that minimizes the magnitude of \(\varepsilon\) is obtained by balancing the exponents (Appendix~\ref{app:error_model_analysis}), which yields
\begin{equation}
    R_c\sim\frac{G^{*}}{2}R_d^2,
\label{eq:optimal_scaling}
\end{equation}
where we have used \(R_d=1/\alpha\).
Let \(G_{\min}\) be the magnitude of the shortest nonzero reciprocal lattice vector,
which equals \(2\pi/h_{\max}\) for the reduced primitive cells used in this work.
If the first active shell coincides with the first geometric shell, \(G^{*}=G_{\min}\), then
\begin{equation}
R_c\sim \pi\,\frac{R_d^2}{h_{\max}},
\end{equation}
which is close to Pickard's empirical form \(R_c\approx 3\,R_d^2/h_{\max}\).
If extinction is present, then \(G^{*}>G_{\min}\) and Pickard's heuristic becomes suboptimal.
This contrast is shown in Figure~\ref{fig:nacl_hcp_scaling}, where NaCl and ideal HCP systems are evaluated with localized Ewald and representative Pickard variants.
For NaCl, \(G^{*}=G_{\min}\), so Pickard's heuristic tracks the derived \(G^{*}\)-aware scaling relation closely.
For ideal HCP, the first geometric shell \((h,k,l)=(0,0,\pm1)\) is extinct, so \(G^{*}>G_{\min}\).
In this case, the \(G^{*}\)-aware scaling relation tracks the optimal scaling path much more closely than Pickard's empirical heuristic.
The same extinction mechanism appears in the more complex Al\(_2\)SiO\(_5\) example, where Figure~\hyperref[fig:al2sio5_analysis]{\ref*{fig:al2sio5_analysis}(a)} shows extinction at \(G_{\min}\), while Figure~\hyperref[fig:al2sio5_analysis]{\ref*{fig:al2sio5_analysis}(b)} shows that the \(G^{*}\)-aware scaling relation follows the optimal scaling path more closely than Pickard's heuristic.
The damped Wolf sum also obeys the optimal scaling relation, as shown in the Supplementary Information.

Furthermore, we observe faster convergence with \(R_c\) of the Pickard variants compared to the localized Ewald method for the ideal HCP and Al\(_2\)SiO\(_5\) systems.
Near the optimal scaling path, the error scales as \(\sqrt{|AB|}\) (Appendix~\ref{app:error_model_analysis}).
The faster convergence thus follows from Pickard's compensation terms reducing the effective prefactor \(A\) in Eq.~\eqref{eq:total_error} relative to localized Ewald.

The above analysis extends naturally to the full Ewald sum.
Let \(G_c^{*}\) be the first active shell with \(G\ge G_c\).
The reciprocal-space truncation error is dominated by this shell, yielding the leading scale \(e^{-(G_c^{*}/(2\alpha))^2}\).
The two-term error model of Eq.~\eqref{eq:total_error} applies with \(G^{*}\) replaced by \(G_c^{*}\), yielding the optimal parameter scaling relation
\begin{equation}
    R_c \sim \frac{G_c^{*}}{2}\,R_d^2
    \quad\Longleftrightarrow\quad
    \alpha \sim \sqrt{\frac{G_c^{*}}{2R_c}},
\end{equation}
where \(G_c^{*}\approx G_c\) when the active shells above \(G_c\) are dense.

\section{Systematic \texorpdfstring{\(G_i^*\)}{G\_i*} Correction}

The omitted reciprocal contributions can also be restored shell by shell.
Let \(\{G_i^*\}_{i=1}^{\infty}\) denote the ordered active shells for which \(C_{\mathrm{shell}}(G_i^*)\) is nonnegligible.
For a generic local electrostatic energy \(E_{\mathrm{loc}}\), we can define a sequence of \(G_i^*\)-corrected energies,
\begin{equation}
    E_{\mathrm{loc}}^{(n)}
    =
    E_{\mathrm{loc}}
    +
    \sum_{i=1}^{n}C_{\mathrm{shell}}(G_i^*)\exp\!\left(-\left(\frac{G_i^*}{2\alpha}\right)^{\!2}\right).
\end{equation}
Each correction removes the contribution of one active shell from the reciprocal-space error.
After adding \(n\) shells, the leading remaining reciprocal contribution shifts to \(G_{n+1}^*\).
The two-term error model [Eq.~\eqref{eq:total_error}] then applies with \(G^*\) replaced by \(G_{n+1}^*\), so the optimal scaling relation shifts accordingly to \(R_c\sim (G_{n+1}^*/2)\,R_d^2\).
Although computing \(C_{\mathrm{shell}}(G_{i}^*)\) requires global positional information, the correction preserves the overall linear-scaling behavior of the real-space methods considered for a fixed number of shells added.

Figure~\ref{fig:al2sio5_analysis} demonstrates the role of these corrections for Al\(_2\)SiO\(_5\), showing that adding more shells accelerates convergence to the Ewald energy for the localized Ewald and Pickard sums.
We also observe that the optimal scaling relation advances from \(R_c=G_1^* R_d^2/2\) in panel~(b) to \(R_c=G_2^* R_d^2/2\) in panel~(c) to \(R_c=G_3^* R_d^2/2\) in panel~(d), as expected.

\begin{figure*}[t]
    \centering
    \includegraphics[width=\textwidth]{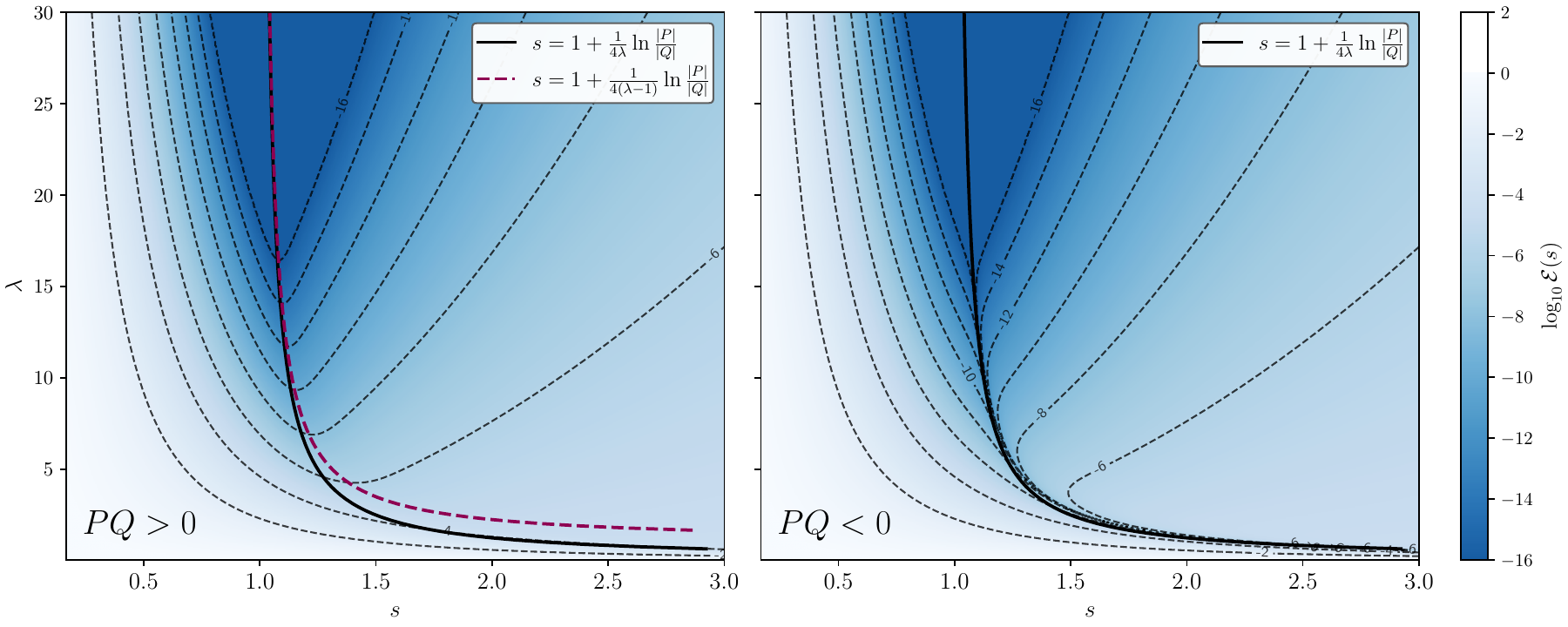}
    \caption{\textbf{Stationarity structure of the error model.}
    Contours of \(\log_{10}\mathcal{E}(s)\) are shown in the \((s,\lambda)\) plane for \(PQ>0\) and \(PQ<0\).
    The \(PQ>0\) panel uses \(P=1\) and \(Q=e^{-5}\), and the \(PQ<0\) panel uses \(P=1\) and \(Q=-e^{-5}\).}
    \label{fig:appendix_stationarity_map}
\end{figure*}

\section{Conclusion}

An error analysis perspective on the real-space electrostatic schemes of Wolf and Pickard has been presented.
Two key analytical results emerge.
First, the analyses in Secs.~\ref{sec:wolf_sum} and~\ref{sec:pickard_sum} demonstrate that the Wolf and Pickard corrections approximate the real-space truncation error in electrostatic sums.
Second, the analysis in Sec.~\ref{sec:error_scaling} presents a nonempirical relation for the optimal scaling path, applicable to a broad class of damped real-space electrostatic methods including the localized Ewald, damped Wolf, and Pickard sums.
These insights have guided the design of a simpler fixed-radii Pickard variant and its force-shifted counterpart, which preserve the compensation behavior of the original Pickard method while improving computational efficiency and enforcing continuity of pair energy and force at the cutoff.
Systematic improvements can be achieved by adding omitted reciprocal contributions shell by shell.

The truncation-error compensation perspective developed here provides a principled basis for designing local approaches to the electrostatic sum.
Furthermore, the results of this work naturally connect to the standard Ewald sum.
A sustained line of work has focused on improving convergence and efficiency of finite-cutoff Ewald evaluations through parameter optimization and explicit correction strategies~\citep{delley1996fast,toukmaji1996ewald,hammonds2022optimization,hammonds2024unification}.
In particular, the \(\alpha'\) method~\citep{hammonds2022optimization,hammonds2024unification}, which compensates for the reciprocal lattice vectors neglected by truncating the reciprocal-space sum, is conceptually complementary to Pickard's corrections, which compensate for the uniform part of the real-space truncation error.
Since Pickard's corrections approximate \(\Delta E_{\mathrm{real\text{-}tail}}\), they could be used as explicit real-space tail corrections to accelerate the convergence of Ewald sums.

\begin{acknowledgments}
We are grateful to Chris Pickard for discussions on the topic.
This work is supported by the Breakthrough Energy Foundation and Robert Bosch LLC.
Claude Code and Codex were used to assist with code implementation, plotting scripts, and mathematical derivations.
\end{acknowledgments}

\appendix

\section{Error Model Analysis}
\label{app:error_model_analysis}

Consider the following error model,
\begin{equation}
    \varepsilon(\alpha)=P e^{-(p\alpha)^2}+Q e^{-(q/\alpha)^2},
\end{equation}
where \(\alpha>0\), \(p,q>0\), and \(P,Q\in\mathbb{R}\).
We define the dimensionless variables
\begin{equation}
    s=\alpha\sqrt{\frac{p}{q}}
    \qquad \text{and} \qquad
    \lambda=pq,
\end{equation}
such that
\begin{equation}
    \varepsilon(s)=P e^{-\lambda s^2}+Q e^{-\lambda/s^2}.
\end{equation}

To minimize the squared error \(\mathcal{E}(s)=\bigl[\varepsilon(s)\bigr]^2\),
we consider its derivative \(\mathcal{E}'(s)=2 \varepsilon(s)\varepsilon'(s)\), where
\begin{equation}
    \varepsilon'(s)=
    -2\lambda s\,P\,e^{-\lambda s^2}
    +
    2\lambda s^{-3}\,Q\,e^{-\lambda/s^2}.
\end{equation}
Stationarity of the squared error requires \(\varepsilon(s)=0\) or \(\varepsilon'(s)=0\),
which reveals two mechanisms for stationary points,
value cancellation (\(\varepsilon=0\)) for \(PQ < 0\), which gives
\begin{equation}
    s^2-s^{-2}
    =
    \frac{L}{\lambda},
\end{equation}
and slope cancellation (\(\varepsilon'=0\)) for \(PQ > 0\), which gives
\begin{equation}
    s^2-s^{-2}
    =
    \frac{L + 4\ln s}{\lambda},
\end{equation}
where \(L=\ln(|P|/|Q|)\).
To obtain the stationary location asymptotically, we consider the limit \(\lambda\to\infty\) and expand in powers of \(\lambda^{-1}\).
The value-cancellation equation can be solved explicitly as a quadratic in \(s^2\) to yield
\begin{equation}
    s=
    \left[
        \frac12\left(
            \frac{L}{\lambda}
            +
            \sqrt{\left(\frac{L}{\lambda}\right)^2+4}
        \right)
    \right]^{1/2}
    =
    1+\frac{L}{4\lambda}+O(\lambda^{-2}).
\end{equation}
For slope cancellation, we consider the ansatz
\begin{equation}
    s = 1 + \delta + O(\delta^2),~\text{for} ~\delta = O(\lambda^{-1}).
\end{equation}
Expanding,
\begin{equation}
    s^2 - s^{-2} = 4\delta + O(\delta^2),
    \qquad
    \ln s = \delta + O(\delta^2).
\end{equation}
Substituting into the stationarity condition and matching terms at leading order gives
\begin{equation}
    4\delta = \frac{L + 4 \delta}{\lambda}
    \quad
    \implies
    \quad
    \delta = \frac{L}{4 (\lambda-1)} + O(\lambda^{-2}),
\end{equation}
so that
\begin{equation}
    s = 1 + \frac{L}{4 (\lambda-1)} + O(\lambda^{-2}).
\end{equation}
In both cases, the leading approximation to the stationary point is \(s=1\), which corresponds to the optimal parameter scaling relation \(\alpha = \sqrt{q/p}\).
These two stationarity mechanisms are shown in Figure~\ref{fig:appendix_stationarity_map}, in the same \((s,\lambda)\) space used for the asymptotic expressions above.
For \(PQ<0\), stationarity is governed by value cancellation \((\varepsilon=0)\), while for \(PQ>0\), stationarity is governed by slope cancellation \((\varepsilon'=0)\).

To characterize how the error near the optimal parameter path scales with \(P\) and \(Q\), we consider the \(\lambda \to \infty\) limit and expand \(\varepsilon\) to first order about \(s^*\approx 1+L/(4\lambda)\),
where each term has magnitude \(\sqrt{|PQ|}\,e^{-\lambda}\), to obtain
\begin{align}
    \varepsilon(s)
    &\approx
    \varepsilon(s^*)+\varepsilon'(s^*)\,(s-s^*),
    \\
    \varepsilon(s^*)
    &\approx
    \bigl[\operatorname{sgn}(P)+\operatorname{sgn}(Q)\bigr]\sqrt{|PQ|}\,e^{-\lambda},
    \\
    \varepsilon'(s^*)
    &\approx
    -2\lambda\bigl[\operatorname{sgn}(P)-\operatorname{sgn}(Q)\bigr]\sqrt{|PQ|}\,e^{-\lambda}.
\end{align}
For \(PQ>0\) (slope cancellation), the slope vanishes and the error at the optimal point is \(2\sqrt{|PQ|}\,e^{-\lambda}\).
For \(PQ<0\) (value cancellation), the stationary point is an exact zero of \(\varepsilon\), and the error grows linearly away from it with slope \(4\lambda\sqrt{|PQ|}\,e^{-\lambda}\).
In both cases, \(\sqrt{|PQ|}\,e^{-\lambda}\) sets the error scale near the optimal path.

\bibliography{references}

\end{document}

\title{Supplementary Information: \\Real-space electrostatic sums through truncation error analysis}

\newcommand\HARVARD{John A. Paulson School of Engineering and Applied Sciences,
Harvard University, Cambridge, MA, USA}

\author{Chuin Wei Tan}
\thanks{Corresponding author}
\thanks{\href{mailto:chuinwei_tan@g.harvard.edu}{chuinwei\_tan@g.harvard.edu}}
\affiliation{\HARVARD}

\author{William C. Witt}
\affiliation{\HARVARD}

\author{Boris Kozinsky}
\thanks{Corresponding author}
\thanks{\href{mailto:bkoz@g.harvard.edu}{bkoz@g.harvard.edu}}
\affiliation{\HARVARD}
\affiliation{Robert Bosch LLC Research and Technology Center, Watertown, MA, USA}

\maketitle

\section{Force-Shifted Pickard Without Shift Correction}
\label{sec:si_pickard_nocorr}

\begin{figure}[h]
\centering
\includegraphics[width=\linewidth]{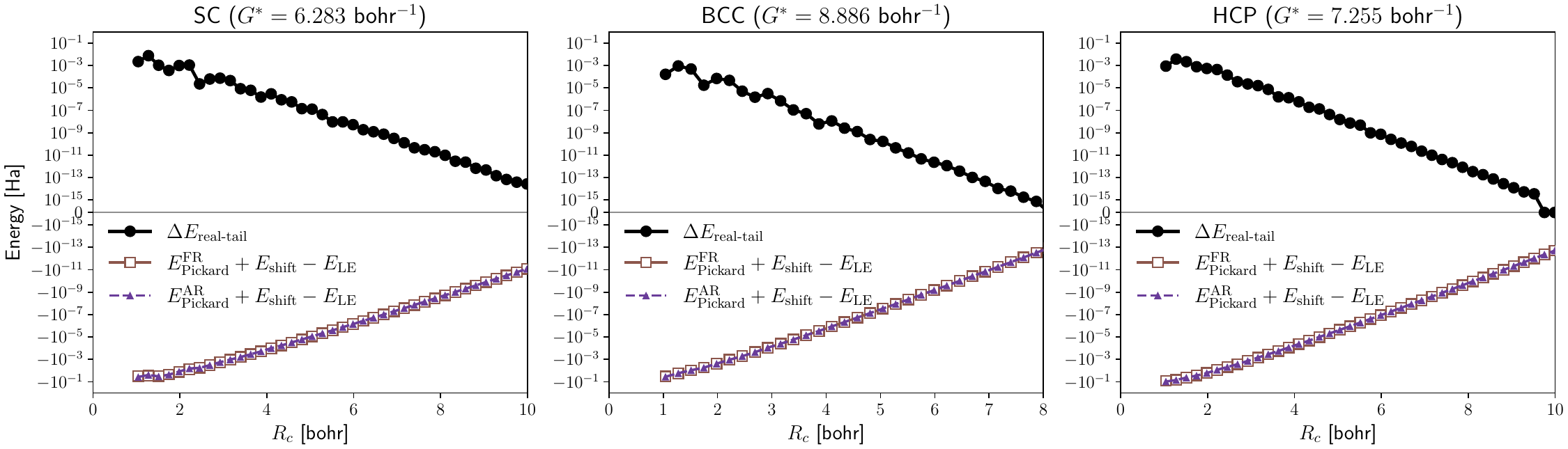}
\caption{\textbf{Truncation-error compensation behavior without the shift correction in force-shifted Pickard variants.}
Results are shown for simple cubic (SC), body-centered cubic (BCC), and ideal hexagonal close-packed (HCP) structures (\(a=1\)~bohr, \(q=+1\) per atom).
The force-shifted adaptive-radii and fixed-radii variants are shown as \(E_{\mathrm{Pickard}}^{\mathrm{AR}}+E_{\mathrm{shift}}-E_{\mathrm{LE}}\) and \(E_{\mathrm{Pickard}}^{\mathrm{FR}}+E_{\mathrm{shift}}-E_{\mathrm{LE}}\), respectively, and compared to \(\Delta E_{\mathrm{real\text{-}tail}}\).
Without the shift correction \(E_{\mathrm{shift\text{-}corr}}\), both force-shifted variants follow the opposite sign relative to \(\Delta E_{\mathrm{real\text{-}tail}}\), indicating loss of truncation-error compensation.
\(R_d\) is scaled with \(R_c\) using \(R_c=G^{*}R_d^2/2\).}
\label{fig:si_pickard_error_analysis}
\end{figure}

\section{Damped Wolf Parameter Maps}
\label{sec:si_wolf_maps}

\begin{figure}[H]
\centering
\includegraphics[width=\linewidth]{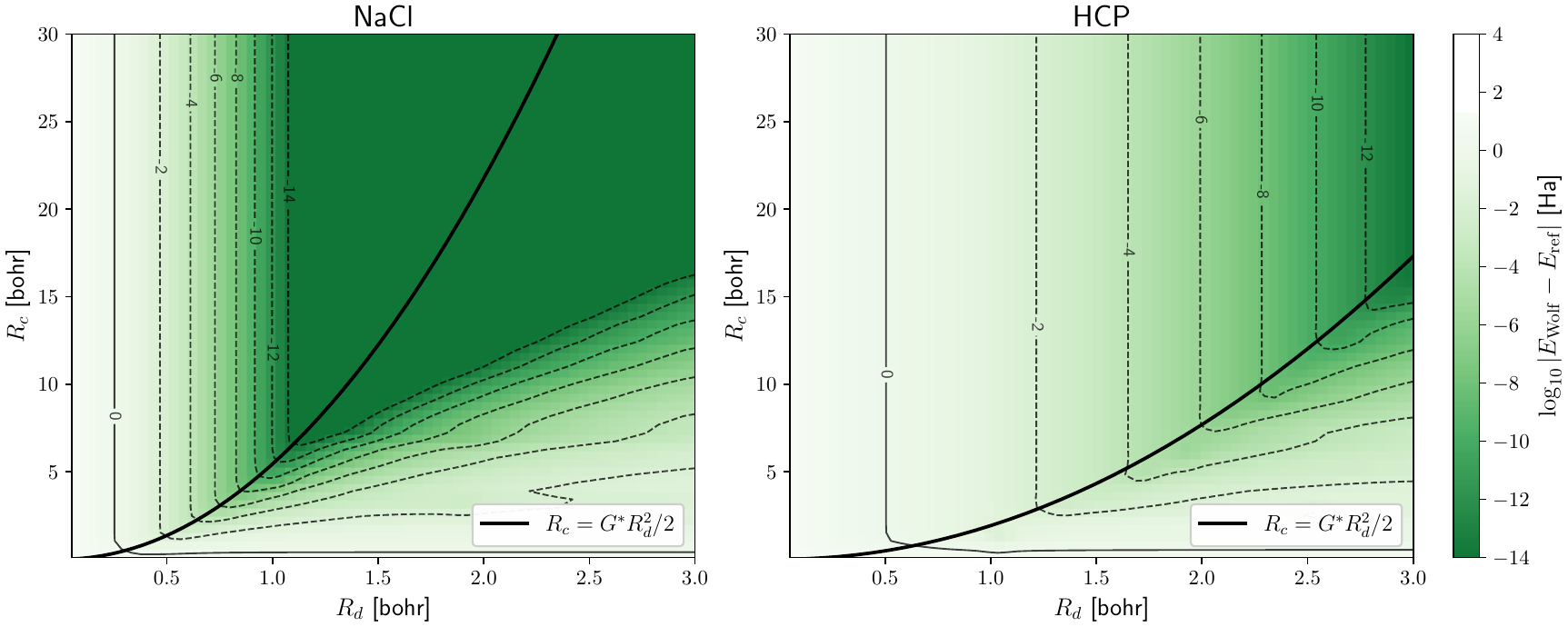}
\caption{\textbf{Damped Wolf convergence maps for NaCl and HCP.}
Error maps of \(\log_{10}|E_{\mathrm{Wolf}}-E_{\mathrm{ref}}|\) in the \((R_d,R_c)\) plane are shown for NaCl (left) and ideal HCP (right) with \(a=1\)~bohr and \(q_1=+1, q_2=-1\),
with reference energies \(E_{\mathrm{ref}}\) from Ref.~\protect\citep{mamode2017computation} for NaCl and a converged Ewald sum for HCP.
The derived scaling relation \(R_c=G^{*}R_d^2/2\) (black) is shown.}
\label{fig:si_wolf_colormap}
\end{figure}

\bibliography{references}